\documentclass[a4paper,10pt]{revtex4-2}
\usepackage{mathrsfs}
\usepackage{graphicx}
\usepackage{latexsym}
\usepackage{amsmath}
\usepackage{amssymb}
\usepackage{textcomp}
\usepackage{amsbsy}
\usepackage{graphics}
\usepackage{epstopdf}
\usepackage{color}

\allowdisplaybreaks[4]

\begin{document}

\tolerance=5000

\title{Singular behavior of the dark universe under the effect of thermal radiation in curved spacetime   }

\author{I.~Brevik,$^{1}$\,\thanks{iver.h.brevik@ntnu.no}
A.~V.~Timoshkin,$^{2,3}$\,\thanks{alex.timosh@rambler.ru}
}
 \affiliation{ $^{1)}$ Department of Energy and Process Engineering,
Norwegian University of Science and Technology, N-7491 Trondheim, Norway\\
$^{2)}$Institute of Scientific Research and Development, Tomsk State Pedagogical University (TSPU),  634061 Tomsk, Russia \\
$^{3)}$ Lab. for Theor. Cosmology, International Centre  of Gravity and Cosmos,  Tomsk State University of Control Systems and Radio Electronics
(TUSUR),   634050 Tomsk, Russia
}

\tolerance=5000

\begin{abstract}

We consider the late-time accelerated universe  in the Friedmann-Robertson-Walker (FRW) spacetime with a nonzero curvature, and investigate cosmological models when  the cosmic fluid is taken to be inhomogeneous and viscous (bulk viscous), coupled to dark matter. We consider the influence from thermal effects caused by Hawking radiation on the  formation of singularities of various classified types, within a finite time. It is shown that under the influence of Hawking radiation the time of formulation of a singularity, and the nature of the singularity itself, can change. It is also shown that by jointly taking into account radiation, viscosity, and space curvature, one can obtain a singularity-free universe.
The symmetry properties of this kind of theory lie in the assumption about spatial isotropy. The spatial isotropy is also reflected in our use of a bulk, instead of a shear, viscosity.

\end{abstract}


\maketitle

\section{Introduction}

As a result of theoretical studies as well as from astronomical observations, the present accelerated expansion of the universe is well established. In particular, the study of the dark energy causing cosmological expansion is of fundamental interest \cite{1,2,3,4}. This kind of energy can be qualitatively described in terms of a cosmological model with negative pressure, satisfying an unusual equation of state \cite{5,6}.

We will consider the universe in the dark energy epoch, because one of the properties of the so-called phantom dark energy (characterized by the thermodynamic parameter  $w$ in the equation of state being less than $-1$) is the possibility to encounter singularities within a finite time $t_{\rm rip}$. According to the Nojiri-Odintsov-Tsujikawa classification of singularities \cite{7,8a} (see also the additional \cite{9a}), one can have the following types, when defining the scale factor $a(t)$, effective (total) energy density $\rho_{\rm eff}$, and effective (total) pressure $p_{\rm eff}$ in the limit $t \rightarrow t_s$,

\begin{itemize}
\item Type I (Big Rip): $a \rightarrow \infty$, $\rho_{\rm eff} \rightarrow \infty$, and $|p_{\rm eff}|\rightarrow \infty$.
\item Type II ("sudden" singularity): $a \rightarrow a_s$, $\rho_{\rm eff} \rightarrow \rho_s$, and $|p_{\rm eff}|\rightarrow \infty$, where $a_s \neq 0$ and $\rho_s$ is constant. This is a pressure singularity.
\item Type III: $a \rightarrow a_s$, $\rho_{\rm eff} \rightarrow \infty$, and $|p_{\rm eff}|\rightarrow \infty$,  This type of singularity is milder than Type I but stronger than Type II.
\item Type IV: $a\rightarrow a_s$, $\rho_{\rm eff} \rightarrow 0$, and $|p_{\rm eff}|\rightarrow 0$, but higher derivatives of the Hubble function diverge. This type also includes the case where $\rho_{\rm eff}$ and/or $|p_{\rm eff}|$ are finite for $t=t_s$.
\end{itemize}
At the same time, it should be recognized that singularities are not the only possible endings of the universe in the phantom phase.

In this paper we will investigate the behavior of the universe near the future singularity, taking into account the joint influence  of bulk viscosity and  Hawking's thermal radiation  on both what  type of singularity that is encountered, and the time of its occurrence. It has been shown earlier that the presence of a viscosity in the cosmic fluid affects the behavior near the singularity \cite{BrevikNojiri2010,BrevikElizalde2011} and thus has to be taken into account when describing the Big Rip type singularity \cite{NojiriOdintsov2003,BambaOdintsov2010,Briscese2007,OdintsovOikonomou2018} and also the singularities of types II, III and IV \cite{OdintsovOikonomou2015,Barrow2013,NojiriOdintsov2004}.

The physical motivation of the presented article is as follows: near the singularity the Hubble function increases, and as a result, an increase of the temperature in the universe occurs. A consequence of this is that  a thermal radiation should appear. Thermal radiation is due to Hawking radiation and is generated at the apparent horizon of the FRW universe \cite{Hawking1975,NojiriOdintsov2020,Capozziello2021,CaiOhta2010}. Hawking radiation is moreover associated with the thermodynamics of black holes, as well as with the existence of a visible horizon of a black hole, and a visible horizon of cosmic events in de Sitter space. The spectrum of thermal radiation is formed at high temperatures in the late universe immediately before the emergence of the singularity. Corrections associated with thermal radiation and with viscosity of the cosmic fluid  near the rip allow us to predict the future state of the universe more accurately.
 The influence of bulk viscosity in the cosmic fluid plays an important role in the Big Rip (BR) singularity, or in the types II, III and IV Rips, at some finite value of time in the future. Recently, the effect of thermal radiation on future singularities of Types I–IV was studied in \cite{Capozziello2021}. It was shown that with singular universes of Types I and III there occurs a qualitative change in the singularity due to thermal effects. The singularities end up as Type II singularities. In universes of Types II and IV, there is no qualitative change in the final state. However, if we take into account a bulk viscosity of the dark fluid, that in the case of radiation, then find, that there is a qualitative change in the singular universe of BR: it may pass into a singularity of Type III, or it may avoid the singularity at all \cite{BrevikTimoshkin2021}. In connection with the foregoing, it is of interest from a physical point of view to study the joint influence of thermal radiation and the properties of viscosity on the type of singularity in space with nonzero curvature. Our universe started from Big Bang which is fundamental singularity. Hence, any study of singularities is of crucial interest.

The aim of this work is to study the influence of thermal radiation at the time of time immediately before the formation of the singularity, taking into account the property of the viscosity of the cosmic fluid for a qualitative change in the singularity of the type of Big Rip in space with nonzero curvature.

At the same time, modern measurements of the luminosity of remote objects by the Planck satellite show that the curvature of the universe is very near zero (apparently less than 0.03 \cite{Planck2019}). Studies related to the distribution of galaxies in space also indicate that the universe is practically flat. However, there are other studies \cite{Valentino2020,Handley2021}  showing  that there is no absolute confidence in this conclusion; it may happen that the universe has a finite curvature after all.

The present work examines the effects of thermal radiation, and bulk viscosity, on the behavior of the late universe when we restrict ourselves to the Big Rip. Emphasis is laid on a qualitative change in the singular behavior in the FRW metric when the curvature is assumed nonzero. We thus obtain a description of a cosmological model induced by an inhomogeneous dark fluid. We discuss how the singular behavior of the universe is associated with thermal radiation, and with the curvature of space.

What role does the concept of symmetry play in this kind of theory? As we are considering the late universe influenced by viscosity and Hawking radiation, there is obviously no symmetry with respect to the early universe, and here is no symmetry like that encountered in bouncing cosmology, for instance. However, the kind of symmetry maintained also in the present case, is that of spatial isotropy as following from the FRW metric. Also, our use of bulk viscosity instead of shear viscosity is in conformity with spatial isotropy. The latter point is actually nontrivial, as in ordinary fluid mechanics the shear viscosity is usually greater than the bulk viscosity.

The aim of this work is to study the influence of thermal radiation at the time of time immediately before the formation of the singularity, taking into account the property of the viscosity of the cosmic fluid for a qualitative change in the singularity of the type of Big Rip in space with nonzero curvature.

\section{Effect from thermal radiation on the formation of singularities in the FRW metric with a nonzero spatial curvature}

We will study the homogeneous and isotropic FRW expanding universe,
\begin{equation}
ds^2 = -dt^2 + a^2(t)\left( \frac{dr^2}{1-\Pi r^2}+r^2 d\Omega^2 \right), \label{1}
\end{equation}
where $d\Omega^2= d\theta^2+\sin^2\theta d\varphi^2$, $t$ is the cosmic time, $a(t)$ denotes the scale factor and has the unit of length, $r$ is the special radius coordinate, and the parameter $\Pi$ is the curvature of the three-dimensional space.

As is known, Eq.~(\ref{1}) geometrically describes different types of the universe. Taking for simplicity $\Pi$ to be nondimensional, for $\Pi = 0$ the universe is spatially flat, for $\Pi =1$ it is closed, and for $\Pi=-1$ it is open. The character of the universe expansion depends on the spatial curvature: the open universe will expand forever,  the flat universe will also expand forever, although at $t \rightarrow +\infty $ the expanding velocity will be constant; the closed universe will expand up to a certain instant, after which the expansion is replaced by a compression leading finally to a collapse.

The Friedmann equation for a one-component fluid in a space with nonzero curvature has the form
\begin{equation}
H^2 = \frac{k^2}{3}\rho_{\rm eff}-\frac{\Pi}{a^2}, \label{2}
\end{equation}
where $\rho_{\rm eff}$ is the effective total energy density, $k^2=8\pi G$ with $G$ the Newtonian gravitational constant, and $H(t)= \dot{a}(t)/a(t)$ the Hubble function.

We will use the following equation of state (EoS) for an inhomogeneous viscous fluid \cite{Capozziello2006},
\begin{equation}
p= \omega(\rho,t)\rho -3H\zeta(H,t), \label{3}
\end{equation}
where $\zeta(H,t)$ is the bulk viscosity, dependent on $H$ and $t$. From ordinary thermodynamics, we know that $\zeta(H,t)>0$.

We take the EoS parameter $\omega$ to have the form \cite{Capozziello2006}
\begin{equation}
\omega(\rho,t)=\omega_1(t)(A_0\rho^{\alpha-1} -1), \label{4}
\end{equation}
where $A_0 \neq 0$ and $\alpha \geq 1$ is a constant. A note on dimensions: as $\omega$ and $\omega_1$ are nondimensional, the dimension of $A_0$ will be complicated when $\alpha >1$. In the simplest case $\alpha =1, A_0$ will be nondimensional. Then we put $\omega(\rho,t)=\omega_0$, a constant.

Dissipative processes are described by the bulk viscosity and  can be given  the form \cite{Capozziello2006}
\begin{equation}
\zeta(H,t)= \zeta_1(t)(3H)^n, \label{5}
\end{equation}
where $n $ is a positive  integer or zero and $\zeta_1(t)$ is an arbitrary function depending on time.

Note that in Eq.~({3}) the dimension of  $\zeta(H,t)$ is cm$^{-3}$ in geometric units, as it should for a viscosity coefficient. Thus $\zeta_1(t)$ in Eq.~(\ref{5}) is for general $n$ not a  viscosity, but a viscosity function. The most important special cases are when $n=0$ and  $n=1$. When $n=0$, $\zeta$ and $\zeta_1$ are equal. When $n=1$, $\zeta(H,t)= 3H\zeta_1(t)$. Thus if $\zeta_1$ is a constant, the bulk viscosity becomes proportional to $H$. The latter special case has attracted considerable attention, as it  appears to satisfy the experimental curves for $H(z)$ versus the redshift $z$ reasonably well. Moreover, as a peculiar property it has turned out that this case allows the universe to pass through  the phantom barrier $\omega =-1$, from the quintessence region $ -1 <\omega <-1/3$ into the phantom region $\omega <-1$ and thus the Big Rip, caused by the bulk viscosity \cite{brevik05}.

 If the fluid is nonviscous, the equation of state reduces to the usual form $p=\omega(\rho,t)\rho$.
 
 \bigskip

We will now consider various models of the viscous fluid, with reference to background material given  in Ref.~\cite{Elizalde2014}.

\subsection{Constant viscosity}

This is the simplest case, where $\omega(\rho,t)=\omega_0$, and $\zeta(H,t)=\zeta_0$, a constant $>0$. The Hubble function takes the form
\begin{equation}
H(t)= \frac{\zeta_0k^2}{1+\omega_0}\,\frac{1}{1-\sqrt{C_1}\exp[\frac{3}{2}\zeta_0k^2t]}, \label{6}
\end{equation}
where $C_1$ is an integration constant. This cosmological model does not take into account the interaction with dark matter.

Using Eq.~(\ref{6}) we calculate the scale factor,
\begin{equation}
a(t)=e^{\int H(t)dt}= a_0\left( 1-\frac{1}{\sqrt {C_1}}\exp[-\frac{3}{2}\zeta_0k^2t]\right)^{\frac{2}{3(1+\omega_0)}}, \label{7}
\end{equation}
where $a_0$ and $C_1$ are constants.

The Hubble function diverges near the singularity time, $t\rightarrow t_s=-\frac{2}{3\zeta_0k^2}\ln \sqrt{C_1}$, meaning that a Big Rip singularity forms. We will study how the type of singularity becomes changed if we take into account thermal radiation, and the curvature of space. The case of a flat space, excluding thermal radiation,  was considered earlier in Ref.~\cite{BrevikTimoshkin2021}.

The physical explanation of this phenomenon is the following. If the temperature in the universe increases near the singularity, thermal radiation is generated. Therewith, the Hubble function becomes large. From statistical physics it is known that the energy density of thermal radiation is proportional to the fourth power of the absolute temperature \cite{NojiriOdintsov2020},
\begin{equation}
\rho_{\rm rad}=\lambda H^4, \label{8}
\end{equation}
 After the thermal radiation component (\ref{8}) is included, the Friedmann equation is modified to become \cite{NojiriOdintsov2020}
 \begin{equation}
 \frac{3}{k^2}H^2 = \rho_{\rm eff}+\lambda H^4. \label{9}
 \end{equation}
 It may be observed that we do not include interacting terms between radiation and matter in Eq.~(\ref{9}), so that the expression may be regarded as an approximate one.
A qualitative inspection of the equation tells us that if the evolution time of the late universe is  much less than the singularity time, the first term on the right-hand side gives the major contribution, while near the singularity time, the last term is most significant. In order to consider the situation more quantitatively, let us solve for the quadratic Hubble function,
\begin{equation}
H^2 = \frac{\frac{3}{k^2} \pm \sqrt{\frac{9}{k^4}-4\lambda \rho_{\rm eff}}}{2\lambda}. \label{10}
\end{equation}
We will apply this result to the cosmological model (\ref{7}). From this, using the Friedmann equation (\ref{2}),  we write $\rho_{\rm eff}$ in terms of the scale factor,
\begin{equation}
\rho_{\rm eff}(t)= \left( \frac{\sqrt{3}\zeta_0k}{1+\omega_0}\right)^2 \left[ \frac{a(t)}{a_0}\right]^{-3(1+\omega_0)}\, e^{-3\zeta_0k^2t}\, \left( \frac{2(\omega_0+1)}{\omega_0+5/3} -e^{-\frac{3}{2}\zeta_0k^2t}\right). \label{11}
\end{equation}
Since $H^2$ is real, the radicand in (\ref{10}) must be non-negative,
\begin{equation}
\frac{9}{k^4}-4\lambda \rho_{\rm eff} \geq 0. \label{12}
\end{equation}
Now assume $C_1=1$. It then follows
\begin{equation}
\frac{9}{k^2}-4\lambda A^2e^{-3\zeta_0k^2t}\left( B-e^{-\frac{3}{2}\zeta_0k^2t}\right) \left[\frac{a(t)}{a_0}\right]^{-3|1+\omega_0|} \geq 0, \label{13}
\end{equation}
where $A= \left( \frac{\sqrt{3}\zeta_0k}{1+\omega_0}\right),$ $B=\frac{2(\omega_0+1)}{\omega_0+5/3}$.

The inequality (\ref{13}) leads to the following restriction on the scale factor,
\begin{equation}
a(t) \leq \left( \frac{2\sqrt{\lambda}Ak^2}{3a_0}\right)^{\frac{2}{3|1+\omega_0|}} e^{\frac{\zeta_0k^2t}{|1+\omega_0|}} \left( B-e^{-\frac{3}{2}\zeta_0k^2t} \right)^{\frac{1}{3|1+\omega_0|}}. \label{14}
\end{equation}
The case of phantom energy corresponds to $\omega_0<-1$. The values of the scale factor are limited by the maximum value $a_{\rm max}$,
\begin{equation}
a(t) \leq a_{\rm max} = a_0\left( \frac{2\sqrt{\lambda}Ak^2}{3}\right) ^{\frac{2}{3|1+\omega_0|}}, \label{15}
\end{equation}
which corresponds to the instant $t_{\rm max}$ given by
\begin{equation}
t_{\rm max}= -\frac{2}{3\zeta_0k^2}\ln \left( 1-\frac{2\sqrt{\lambda}Ak^2}{3}\right). \label{16}
\end{equation}
Note that the new time of formulation of singularity coincides with that obtained for a universe with flat metric.

We can calculate the difference between $t_{\rm max}$ and $t_s$,
\begin{equation}
t_{\rm max}-t_s = -\frac{2}{3\zeta_0k^2}\ln \left( 1-\frac{2\sqrt{\lambda}Ak^2}{3}\right) > 0, \label{17}
\end{equation}
which shows that $t_{\rm max}$ is larger than $t_s$. The time of singularity changes: it occurs later.

To determine which type of singularity that occurs, we calculate the effective energy density $\rho_{\rm eff}$ and the effective pressure $p_{\rm eff}$ when $t\rightarrow t_{\rm max}$. In this limit $\rho_{\rm eff} \rightarrow \rho_{\rm max}$ and $|p_{\rm eff}|\rightarrow |p_{\rm max}|$, and we get
\begin{equation}
\rho_{\rm eff}(t_{\rm max})= 3\left(  \frac{3\zeta_0k}{1+\omega_0}\right)^2 \left( \frac{3\omega_0+1}{(2\sqrt{\lambda}Ak^2)(\omega_0+5/3)}+1 \right)^2, \label{18}
\end{equation}
\begin{equation}
 |p_{\rm eff}(t_{\rm max})|= |\omega_0\rho_{\rm max}-3\zeta_0H_{\rm max}|= \frac{3\zeta_0^2k^2}{|1+\omega_0|} \left[ \frac{9\omega_0}{|1+\omega_0|}
\left( \frac{3\omega_0+1}{2\sqrt{\lambda}Ak^2(\omega_0+5/3)}+1\right) -1+ \frac{3}{2\sqrt{\lambda}Ak^2}\right]. \label{19}
\end{equation}
Thus, the values of the scale factor, energy density, and effective pressure are finite, while the high derivatives of $H$ do not diverge. The singularity encountered is accordingly of type IV. The result can be physically explained by the influence from the viscosity of the dark fluid. It compensates for the effects of thermal radiation and space curvature.

\subsection{Viscosity proportional to the Hubble function}

In this subsection we assume the bulk viscosity to be proportional to the Hubble function, $\zeta(H,t)=3\tau H$, where the constant $\tau$ is positive. [In natural units where the fundamental length is cm, the dimension of $\zeta$ is cm$^{-3}$, and since the dimension of $H$ is cm$^{-1}$, the dimension of $\tau$ is cm$^{-2}$.]

The Hubble function takes the form
\begin{equation}
H(t)= \frac{k}{\sqrt{3}}\,\frac{\delta}{3\theta +\exp[-\frac{1}{2}\tilde{\eta}t]}, \label{20}
\end{equation}
where $\delta$ is a positive constant, $\tilde{\eta}=\delta \tilde{\gamma}, \,\tilde{\gamma}= \frac{k}{\sqrt{3}}\sqrt{1+r},  \, \theta = 1+\omega_0 -9 \tau {\tilde{\gamma}}^2. $ The constant $r$ is associated with the influence from dark matter,
and is defined as the ratio of the energy density of dark matter to the energy density of dark energy. In the case where $\omega_0 < -1+9\tau {\tilde{\gamma}}^2$, we have $\theta <0$ and the Hubble function diverges at $t\rightarrow t_s = -\frac{2}{\tilde{\eta}}\ln (-3\theta).$ A singularity of type Big Rip is formed.

Let us now see how the type of singularity may change. First, calculate the scale factor
\begin{equation}
a(t)= a_0e^{\alpha t}\left( e^{-\frac{1}{2}\tilde{\eta}t}+3\theta\right)^{2\alpha/\tilde \eta}, \label{21}
\end{equation}
where $\alpha = \frac{\delta \gamma}{3\sqrt{3}\,\theta} <0$ with dimension cm$^{-1}$.

The energy density expression, in terms of the scale factor, gets the form
\begin{equation}
\rho_{\rm eff}=\left( \frac{\delta \gamma}{k}\right)^2 e^{\tilde{\eta}t}
\left[ \frac{a(t)}{a_0}\right]^{\frac{\tilde{\eta}}{|\alpha|}}
 \left[ 1+\frac{3(3\tau k^2-\omega_0-1)}{3\omega_0+1} -
\sqrt{3}\, e^{-\frac{1}{2}\tilde{\eta}t} \frac{3\theta +e^{-\frac{1}{2}\tilde{\eta}t}}{k\delta \gamma(3\omega_0+1)}\right]. \label{22}
\end{equation}
If $|\alpha|= \tilde{\eta}$, we obtain the following restriction on the scale factor:
\begin{equation}
a(t) \leq 3a_0|3\omega_0+1|(A+B|a|)C^2e^{-|\alpha| t}, \label{23}
\end{equation}{
where $A= 2-9\tau k^2, \, B= \frac{\sqrt{3}}{\delta\gamma}, \, C= \frac{3}{2\sqrt{\lambda} \,\gamma \delta k}.$}

Hence the scale factor is restricted by the maximum value $a_{\rm max}$,
\begin{equation}
a(t) \leq a_{\rm max}=3a_0|3\omega_0+1|(A+B|\alpha|)C^2. \label{24}
\end{equation}
It corresponds to the instant
\begin{equation}
t_{\rm max}= -\frac{2}{|\alpha|}\ln \frac{1}{2}T\left( 1+\sqrt{1-\frac{12|\theta|}{T}}\right), \label{25}
\end{equation}
where $T= C\sqrt{|3\omega_0+1|(A+B|\alpha|)}$, and $T<\frac{1}{1-3|\theta|}$.

Let us find the difference between the singularity times. If we for simplicity put $T= 12|\theta|$, then
\begin{equation}
t_{\rm max}-t_s= \frac{2}{|\alpha|}\ln \frac{1}{2}<0, \label{26}
\end{equation}
which shows that $t_s$ is larger than $t_{\rm max}$. Then in the limit $t\rightarrow t_{\rm max}$ the effective energy density $\rho_{\rm eff} \rightarrow \rho_{\rm max}$ and effective pressure $|p_{\rm eff}| \rightarrow |p_{\rm max}|$ become finite,
\begin{equation}
\rho_{\rm max}= \frac{3}{|3\omega_0+1|} \left( \frac{\delta \gamma}{k}\right)^2\left( 2+\frac{2-9\tau k^2}{3|\theta|}\right), \label{27}
\end{equation}
\begin{equation}
|p_{\rm max}|= |\omega_0\rho_{\rm max}-9\theta H_{\rm max}^2| = \left( \frac{\delta\gamma}{k}\right)^2 \Big|\frac{\omega_0}{3\omega_0+1}
\left( 2+\frac{2-9\tau k^2}{3|\theta|}\right)
-\frac{3\tau k^2}{(3\theta +T/2)^2}\Big|. \label{28}
\end{equation}
However, the higher derivatives of the Hubble function do not diverge. Thus, formation of a type IV singularity does never occur in this model.

\section{Conclusion}

As recently shown in \cite{NojiriOdintsov2020}, for an ideal fluid near the singularity it is necessary to take into account the Hawking thermal radiation, which leads to a change in this  singularity's type.  For singularities of type I or III with a finite time of formation a transition to a type II singularity occurs, so there is even  a qualitative change taking place. For singularities of type II or IV, there occurs however no qualitative change in the final state.

Following our earlier paper \cite{BrevikTimoshkin2021} we took into account a bulk viscosity of the dark fluid, and a modified equation of state in the case of thermal radiation. We showed that there is a qualitative change in the behavior of a singular universe of type I: it may pass into a singularity of type III, or it may avoid the singularity at all in a finite time. This absence of singularity was explained by the fact that the effect of thermal radiation  becomes  neutralized by the viscosity in the fluid.

As a summary of the present investigation, we have analyzed thermal effects due to the Hawking radiation on the singular behavior in the dissipative model of the dark universe in the Friedmann-Robertson-Walker metric with a nonzero spatial curvature. We showed that a combined theory accounting simultaneously for radiation, viscosity, and curvature, may lead to the absence of singularities at all.

The novelty of the article is that in the space with nonzero curvature singularity, they are not formed in contrast to the results of \cite{BrevikTimoshkin2021} in the flat space, where, singular type III are possible.

It may also be mentioned that the inclusion of thermal radiation in the theory leads to good agreement with astronomical observations \cite{Astashenok2021}.

\end{document}